\documentclass[a4paper, 12pt]{article}
\usepackage[english]{babel}
\usepackage[a4paper, inner=2cm, outer=2cm, top=3cm, bottom=3cm]{geometry}
\usepackage{mathtools}
\usepackage{pstricks}
\usepackage{caption}
\usepackage{graphicx}
\usepackage{amsmath}
\usepackage{amssymb}
\usepackage{mathcomp}
\usepackage{textcomp}
\usepackage[doublespacing]{setspace}
\usepackage{etoolbox}
\usepackage{romannum}
\usepackage{graphicx}
\usepackage{caption}
\usepackage{subcaption}

\title{On the quantum improved Schwarzschild black hole}
\author{R. Moti and A. Shojai\\
\textit{\small Department of Physics, University of Tehran, Tehran, Iran}}
\date{}
\begin{document}
\maketitle
\pagenumbering{arabic}
\begin{abstract}

Deriving the gravitational effective action directly from exact renormalization group is very complicated, if not impossible. Hence, to study the effects of running gravitational coupling which tends to a non--Gaussian UV fixed point (as it is supposed by the asymptotic safety conjecture), two steps are usually adopted. Cutoff identification and improvement of the gravitational coupling to the running one.

 As suggested in \cite{Moti-Shojai3}, a function of all independent curvature invariants seems to be the best choice for cutoff identification of gravitational quantum fluctuations in curved spacetime and makes the action improvement, which saves the general covariance of theory, possible. Here, we choose Ricci tensor square for this purpose and then the equation of motion of improved gravitational action and its spherically symmetric vacuum solution are obtained. Indeed, its effect on the  massive particles' trajectory and the black hole thermodynamics are studied.

\end{abstract}
\section{Introduction}
Unfortunately among all fundamental interactions, gravity (described best by General Relativity) when quantized, blurts out offbeat behaviors such as divergences in the description of high energy phenomena.
Although the unusual properties of gravity outcrop at energy scales for which other three interactions are unified by a well--known quantization formalism, its long range action cannot be ignored at macroscopic scales, i.e. below Planck mass $m_{Pl}$.
Hence, any quantization method for gravity at high energies affects on the physics at scales for which we have a fine description for other interactions.

Although the undesirable divergences of quantization of Einstein theory may caused from the usual perturbative approach, it is noteworthy that there are debates that quantum mechanics and general relativity are compatible, because of their different viewpoints about some fundamental concepts like time (see references in \cite{Time}).

Assuming that we can find compatible quantization of general relativity, various approaches in dealing with this challenge are suggested \cite{Rovelli}. 
However, with all of the efforts, a covariant renormalizable approach which is faithful both to the particle physics and general relativity principles is not suggested yet.
Use of modern renormalization methods to renormalize the divergences of gravitational field theory is one of the considerable suggestions.

As Weinberg \cite{Weinberg-1st,Weinberg-2nd} suggested it may be possible to consider gravity as an asymptotically safe theory at high energies. In other words, the dimensionless gravity coupling may run on the trajectory to a \textit{non--Gaussian} fixed point which is located on a finite dimensional surface on the Planck mass energy scale and stays safe from divergences.
This assumption in addition to considering a finite number of essential couplings (those which cannot be eliminated by field redefinition), lead to a predictive renormalizable quantum gravity theory, which attracts many attentions \cite{Attentions}.

Various methods in probing the existence of non--Gaussian fixed points are invented (see references in \cite{Weinberg inflation}). 
Using the functional renormalization group, Reuter defined the gravitaional flow which its scale dependence is determined by the exact renormalization group equation (ERGE) \cite{Reuter-1st}. 
Hence the running gravitational coupling which has a fixed point on its trajectory to UV regime is derived \cite{Souma}.

Such effective couplings can affect many solutions of general relativity, including cosmology \cite{Weinberg inflation, Moti-Shojai} and black holes \cite{Reuter Becker, Falls, Bonanno & Reuter, Pawlowski}.
Although introducing variable couplings can be found in other theories such as scalar--tensor theories (see references in \cite{Varing G}),
but here the dynamics of the gravitational constant is a result of quantum running which comes from ERGE, not as an input or as a dynamical field.

One of the best places to investigate the validity and results of quantum gravity is black holes, hence the effect of RG improvement of couplings on the behavior of black hole singularities of spacetime is remarkable. 

This is partially investigated in the literature \cite{Reuter & Weyer & Noorbala}, but here, we probe this effect on the Schwarzschild black hole using the method called \textit{action improvement} \cite{Moti-Shojai3}. In the next section, after a short review on ERGE, we introduce the method briefly. Then we will discuss its results and properties on the Schwarzschild black hole in section \ref{3}. The last section is devoted to the comparison of the results of our method of action improvement, and the existing ones in the literature.

\section{Quantum improved gravity} \label{2}
In order to see the effect of asymptotic safe gravity, the first step is finding a suitable gravity flow $\Gamma_{k}[g_{\alpha\beta}]$ which is the solution of the evolution equation ERGE \cite{Wetterich}
\begin{equation}
  \frac{\partial}{\partial k} \Gamma_{k}[\phi] = \frac{1}{2} Tr{[\Gamma_{k}^{(2)}[\phi]+ {\cal R}_{k}]^{-1} \frac{\partial}{\partial k}{\cal R}_{k}} \ .
\end{equation}
The $\Gamma_{k}[g_{\alpha\beta}]$, can be considered as an effective average action for the gravitational field and $\mathcal{R}_k(p^2) \propto k^2 \mathcal{R}^{(0)}(p^2/k^2) $ is the IR cutoff term.
This arbitrary smooth function is the result of adding IR cutoff term to the classical effective action to suppress the low momentum modes, where $ \mathcal{R}^{(0)}(\psi) $ satisfies the conditions $ \mathcal{R}^{(0)}(0) =1 $ and $ \mathcal{R}^{(0)}(\psi \rightarrow \infty) \rightarrow 0$.
It is required that $\mathcal{R}_{k\rightarrow0}$ vanishes in order to not disturb high momentum modes.
Usually, an exponential form $ \mathcal{R}^{(0)}(\psi) = \psi/(\exp(\psi) -1) $ is used in the literature \cite{Nagy}.

Because of the fact that considering all terms of this generating functional which interpolates $\Gamma_{k\rightarrow \infty} = S$ (any admissible fundamental action) to $\Gamma_{k\rightarrow 0} = \Gamma$, changes the ERGE to an unsolvable one by the known mathematical methods, one may use a \textit{truncation} which restricts RG flow to a finite dimensional subspace.
The Einstein--Hilbert truncation 
\begin{equation}\label{eq2.1}
 \Gamma_k[g_{\alpha\beta}] = \frac{1}{16\pi G_{k}}\int{d^{4}x\sqrt{g}(-R(g)+2\Lambda_{k})} + S_{gf}[g_{\alpha\beta}] \ ,
\end{equation}
projects $\Gamma_{k}[g_{\alpha\beta}]$ on the subspace which is spanned by $\int\sqrt{g}R $ and $\int\sqrt{g}$ and
seems to be a suitable finite--dimensional subspace for studying the effect of this method on GR. $S_{gf}[g_{\alpha\beta}]$ is the gauge fixing term.
Since we probe the Schwarzschild solution, the approximation $ \Lambda_{k} \approx 0 $ is chosen in what follows.

On using this truncated $\Gamma_{k}[g_{\alpha\beta}]$ in the ERGE, one can find \cite{Bonanno & Reuter} the $\beta$--function, which defines the evolution of the couplings.
The analytical solution of this $\beta$--function at \textit{infrared regime} ($k\rightarrow0$) is $  G(k) = G_0 \left[ 1 - \omega G_0 k^2 + \mathcal{O}(G_0^2 k^4)\right]$, while near the \textit{fixed point} ($k\gg m_{pl}$), it is $G(k) = g_{*}^{UV}/{k^2}$. These two analytical solutions can be combined and written as
\begin{equation}
  G(k) = \frac{G(k_{0})}{1 + \omega G(k_{0})(k^{2}-k_{0}^{2})}
\end{equation}
where $ k_{0} $ is a reference scale, with the condition $ G_{N} \equiv G(k_{0} \to 0)=G_0$
in which  $G_N$ is the experimentally observed value of Newton's constant, and $\omega = \frac{4}{\pi}(1-\frac{\pi^{2}}{144})$ \cite{Bonanno & Reuter}.

After obtaining the running coupling constant, one has to identify the cutoff scale. In the same way as applying Uehling correction to Coulomb potential in massless QED, the renormalization momentum $k$  should be identified with a single dimensionful parameter of the problem, $r$ \cite{Dittrich & Reuter}.
Dimensional analysis suggest a general identification $k(r) = \xi/D(r) $ where $D(r)$ is a distance function and $\xi$ is a dimensionless parameter.
In contrary to QED, the distance scale does not have a unique definition in general relativity and is a debatable issue in the viewpoint of general covariance.
Although for the asymptotic flat region,  $r\gg r_{s}$ (where $r_s=2G_NM_s$) in the Schwarzschild black hole, the choice $D(r)=r$  is well defined, but in curved regions various distance functions could be defined.
It is usual to use for the whole spacetime, the simple cutoff identification $D(r)=r$ and $\xi\simeq 1$ which is quite reasonable for large $r$. This cutoff identification leads to $r$-dependent of gravity coupling
\begin{equation} 
    G(r) = \frac{G_N}{ \left( 1+ \omega G_{N}/r^2 \right)} \ .
    \label{G-r}
\end{equation}
For more cutoff identifications see \cite{Nagy}, \cite{Bonanno & Reuter} and references in \cite{Falls}.

The next step is to use this running coupling constant (obtained by solving the ERGE)  to improve the gravity theory, such that the quantum effects are included. The way we use \textit{variable} fundamental constants is a questionable topic \cite{Ellis}, and thus the decision of where and how to exert the improvement of $G_{N}$ to $G(r)$, is the final step in this method. Different ways of improvement can be categorized as follows \cite{Moti-Shojai3}:

\begin{itemize}
\item \textit{Solution improvement}: In this proposal, $G_N$ is replaced with the obtained running coupling constant $G(x)$ as an input function, in the non--improved solutions of Einstein's equations. Clearly this is the simplest way and most debatable way of improving the classical results.

\item \textit{Equation of motion improvement}: In this second way of improvement,  the replacement of $G_{N}$ with $G(x)$ is done at the level of the equation of motion and not the solution. That is to say, the Einstein's equations would be $G_{\mu\nu}=-8\pi G(x) T_{\mu\nu}$.  It is clear that for vacuum solutions, equation of motion and solution improvement are equivalent. Notice that the presence of the gravitational constant in a vacuum solution like the Schwarzschild back hole is through the Newtonian limiting case.
The differences of these two methods become bold for  non--vacuum solutions and the latter one seems to be more acceptable.

\item \textit{Parameter improvement}\footnote{Although this method is called action improvement by some authors \cite{Reuter & Weyer & Noorbala}, we believe that this name is more appropriate  for a different way of improvement we propose as the next method.}: In the parameter improvement,  one replaces  $G_N$ with $G(x)$ in the Einstein--Hilbert action, without adding any kinetic term for it. Then the gravitational equation of motion are obtained from this new action with externally prescribed field $G(x)$. Note that since $G(x)$ is obtained in some specific frame of reference, we shall loss the general covariance.

In order to make this method physically more acceptable, one should add kinetic terms for $G(x)$ in such a way that the resulting equation of motion for $G(x)$ has exactly the solution obtained from the ERGE. This is the consistency condition. The bad news for this way of improvement is that this is a hard task to do.

\item \textit{Action improvement}: As it is clear, none of the above mentioned methods of improvement are physically plausible.
In a physically acceptable improvement method, the quantum corrections have to lead to a general covariant improved action. The improved solutions are the solutions of the improved Einstein's equations obtained from the improved action.

To do so, here we suggest to improve $G_N$ to $G(\chi )$ in the action, where $\chi$ is a function of all independent curvature invariants. That is to say, the proposal is to use the cutoff identification $k\to\chi$.
This is a true physical cuttoff identification, because as discussed in the \cite{Moti-Shojai3}, the suitable length scales for this method are given by the curvature in general relativity.
Since the tidal forces are described by the curvature components, we could consider this cutoff as the maximum neighbourhood size which can be considered so small that the tidal forces of gravitational quantum fluctuations cannot violate the equivalence principle yet \cite{Moti-Shojai3}.

To obtain $G(\chi)$, first we should find  $x(\chi)$ from the non--improved solution and then $G(\chi)$ is defined.
By this substitution the general covariance would be saved without any critique for ignorance of kinetic terms of the field $G(x)$. 

It has to be noted that in this way we are dealing with a higher derivative theory.
\end{itemize}

In what follows we implement the action improvement for Schwarzschild solution and investigate the result of such an improvement on this black hole.

\section{Action improved Schwarzschild black hole}\label{3}
Schwarzschild black hole is the most known vacuum solution of Einstein's equations. As it is stated before, since it is a vacuum solution solution and equation of motion improvements would have identical results. The emergence of the gravitational constant, $G_{N}$, comes from applying weak field limit.
The solution improved metric components are thus $g_{00} = - g_{11}^{-1} = 1 - \frac{r_{s}}{r} + \frac{\omega G_{N}r_{s}}{r^3} $.  There are features of this improved metric investigated in the literature.
As mentioned in \cite{Falls}, the improved temperature  is lower than the non--improved one. It would have negative specific heat and the black hole evaporates through the Hawking emission. For this improved solution, one can obtain no, one or more horizons depending on the improvement parameter $\omega$.

As we discussed in the previous section, action improvement can be considered as a better way of improvement, and thus it is natural to look for the action improved black holes and their physical properties. In the following first we find the improved action, and then obtain the corresponding equation of motion for spherical symmetry.  Equation of motion is solved the effects on the vacuum static spherical symmetric solutions is studied.
\subsection{Improved action and equation of motion}
We saw that to have a well--defined action improvement, and to save the general covariance, we should change the dependence of running coupling on the cutoff identification parameter (say, $r$) to a function of the curvature invariants.

In what follows, we choose a simple one $\chi\equiv R_{\mu\nu} R^{\mu\nu}$.
Although, as it is discussed in the previous section, in general $\chi$ could be a function of all the independent curvature invariants, but 
an important question is that what is this function? It should be determined via some physical conditions and/or selection rules.
This could be energy conditions, and the behavior of the geodesic congruence as discussed in \cite{Moti-Shojai4}, or other conditions as in a forthcoming work. For simplicity and comparison of the results with other improvement methods, the choice $\chi= R_{\mu\nu} R^{\mu\nu}$ is chosen here.

As a result, action can be improved to
\begin{equation} \label{eq4.3}
  S = \frac{1}{16 \pi} \int d^4x \frac{\sqrt{-g}}{G(\chi)}R \ . 
\end{equation}
The least action principle results in the following modified equations of motion
\begin{equation} \label{eq4.4}
  G_{\mu\nu} = X_{\mu\nu} / \mathcal{J}
 \end{equation}
 with
 \[ X_{\mu\nu} = 
 \]
 \begin{equation} \label{eq4.6}
      \mathcal{K} R R_{\mu\alpha} R_{\nu\beta} g^{\alpha\beta} + \nabla_{\mu}\nabla_{\nu}\mathcal{J} - \nabla^{\sigma}\nabla_{\nu}(  \mathcal{K} R R_{\mu\sigma}) 
   -g_{\mu\nu} \square \mathcal{J} + \frac{1}{2} g_{\mu\nu} \nabla_{\rho}\nabla_{\sigma} ( \mathcal{K} R R^{\rho\sigma})+\frac{1}{2}\square( \mathcal{K} R R_{\mu\nu})
\end{equation}
where $\mathcal{J} \equiv G^{-1}(\chi) $ and $ \mathcal{K} \equiv \frac{2 \partial G/\partial\chi}{G(\chi)^2} $.

In order to make the model complete, we have to obtain the form of $G(\chi)$. To do so, one strategy may be the investigation of the asymptotic behavior of $\chi$ and $G$. Let's first consider the asymptotic behavior of $\chi$. For this purpose, we use the static spherically symmetric spacetime

\begin{equation} \label{eq4.1}
  ds^2 = f(r) dt^2 - g(r) dr^2 -r^2 d\theta^2 - r^2 \sin^2\theta d\varphi^2 \ .
\end{equation}

For the asymptotic flat region, we set $f(r) = 1+u(r) $ and $g(r) = 1+v(r)$, with $v(r)=-u(r)\ll 1$. The components of Ricci tensor at this approximation are
\begin{equation}
  R_{00} \simeq \frac{v^{\prime\prime}}{2} +\frac{v^{\prime}}{r}, \quad R_{11} \simeq -\frac{v^{\prime\prime}}{2} +\frac{u^{\prime}}{r}, \quad R_{22} \simeq \frac{1}{r^4}(u+\frac{r}{2}(u^{\prime}+v^{\prime})), \quad R_{33} \simeq \frac{1}{r^4 \sin^2\theta}(u+\frac{r}{2}(u^{\prime}+v^{\prime})) \ .
\end{equation}
Hence, at the asymptotic region $v(r)\ll 1$, the dependence of $ r $ on the scalar invariant $\chi $ is the solution of the relation
$ \chi = 2(\frac{v^{\prime\prime}}{2} +\frac{v^{\prime}}{r})^2 + \frac{2}{r^4} v^2 $.
Using the non--improved solution $v(r)\sim r_s/r$, we get $\chi^2\sim 10r_s^2/r^6$.

Analogous to QED \cite{Reuter-1st}, one of the customary cutoff identifications in the asymptotic flat region is $k\varpropto r^{-1}$. Using the relation (\ref{G-r}) we finally get the following relation for spherically symmetric spacetime
\begin{equation}
G(\chi)=\frac{G_N}{1+\omega G_N\left (\dfrac{\chi^2}{10r_s^2}\right)^{1/3}} \ .
\label{G-chi}
\end{equation}
Although this is obtained for the asymptotic flat region, we \textit{extend} it to all the spacetime regions. This equation when extended to the whole spacetime, is clearly covariant and so the action (\ref{eq4.3}) and the equations of motion (\ref{eq4.4}). 

Equations (\ref{eq4.4}) and (\ref{G-chi}) form a set of complete and closed equations for obtaining the improved solution.

It is interesting to note that one can rewrite the action with this relation for $G(\chi)$, using the method of Lagrange multiplier as
\begin{equation}
  S = \frac{1}{16 \pi} \int d^4x \sqrt{-g} \left (\frac{R}{G(\chi)}+\lambda \left [ G(\chi)-\frac{G_N}{1+\omega\zeta \chi^{2/3}} \right ]^2 \right )
\end{equation}
where $\lambda$ is the Lagrange multiplier and $\zeta$ is a constant.
\subsection{The solution} 
Solving equation \eqref{eq4.4} exactly, even for spherically symmetric spacetime is almost impossible.  But since the improvement effects are quantum corrections to the classical non--improved solution, one can expand the solution around the  Schwarzschild metric $ f(r) = 1 - r_{s}/r$ and $ g(r) = -(1-r_{s}/r)^{-1}$.
This means that we can write
\begin{equation} \label{eq4.7}
     g_{00} = 1 - \frac{r_s}{r} + l(r) , \quad
     g_{11} = -(1 - \frac{r_s}{r})^{-1} + p(r), \quad 
     g_{22} = -r^2 , \quad
     g_{33} = -r^2 \sin^2\theta
\end{equation}
and solve the equations of motion perturbatively.
Note that $ X_{\mu\nu} $ doesn't have zeroth order term, and thus to solve the equations of motion up to the first order, we just have to consider zeroth   order term of ${\cal J}^{-1}$ factor (i.e. $G_N$) in the right hand side of equation \eqref{eq4.4}.
Substituting the above metric components in \eqref{eq4.4} and after some lengthy calculations, we get the following equations, up to the first order:
\begin{align} \label{eq4.8}
  & p^{\prime} + p \frac{1}{r} \frac{1+r_s/r}{1-r_s/r} = -\Omega_{M_{s}}\frac{(r_s/r)^2}{r(1-r_s/r)}\left(1+\frac{r_s/r}{2(1-r_s/r)} \right) \ , \nonumber \\
  & l^{\prime} - l \frac{1}{r} \frac{r_s/r}{1-r_s/r} + p \frac{1}{r}(1-\frac{r_s}{r}) = -\Omega_{M_{s}}\frac{(r_s/r)^2}{r}(\frac{5r_s}{2r}-2) \ ,\\
  & \begin{multlined}[t]l^{\prime\prime} \frac{r^2}{2} +l^{\prime} \frac{r}{2}(1-\frac{r_s/r}{2(1-r_s/r)}) + l\frac{r_s}{4r} (\frac{1}{1-r_s/r} +\frac{1}{(1-r_s/r)^2} ) + \\
  p^{\prime} \frac{r}{4} (1-\frac{r_s}{r}) (2-\frac{r_s}{r}) +  p \frac{r_s}{4r}(2-\frac{r_s}{r}) = -2\Omega_{M_{s}} \frac{r_{s}^{2}}{r^2} (1-\frac{r_s}{r}) \end{multlined} \nonumber
\end{align}
where $ \Omega_{M_{s}} \equiv  \frac{2\omega G_{N}}{r_{s}^{2}}=2\omega \left(\frac{\ell_{Pl}}{r_s}\right)^2$ and prime denotes differentiation with respect to $r$.
The first two equations  are coupled first order differential equations.  They can be simply solved to get
\begin{align}
  & g_{00} = f(r) = 1-\frac{r_{s}}{r} + l(r) = (1-\frac{r_{s}}{r})\left(1+\Omega_{M_{s}}(A-\frac{2\frac{r}{r_{s}}(2B\frac{r}{r_{s}}+1)-5}{4\frac{r^2}{r_{s}^2}(\frac{r}{r_{s}}-1)})\right) \label{eq4.9} \ , \\
  & g_{11} = g(r)=-(1-\frac{r_{s}}{r})^{-1} + p(r) =-(1-\frac{r_{s}}{r})^{-1}\left(1+\Omega_{M_{s}}\frac{1+4\frac{r}{r_s}(B\frac{r}{r_s}-1)}{4\frac{r^2}{r_s^2}(\frac{r}{r_s}-1)}\right) \label{eq4.10}
\end{align}
where $A$ and $B$ are integration constants. It has to be noted that to save the correct signature of the metric, the constant $A$ has to be restricted by the relation $ \Omega_{M_{s}} A > -1+\frac{3 r_s^2}{8 r^2}\Omega_{M_{s}} $.

These solutions should be checked that if are compatible with the third equation of (\ref{eq4.8}). Putting the above solution in it, one sees that it is valid up to our approximation $\mathcal{O}(\Omega_{M_{s}}r_{s}^2/r^2)$. 

It should be noted that because of the cutoff identification used to relate the momentum scale to the length scale, we are prevented from using this solution near UV fixed point where $k \rightarrow \infty $ or $r \rightarrow 0$. 

Near the UV fixed point we have to go back to equations (\ref{eq4.4}) and (\ref{eq4.6}), and on using (\ref{eq4.7}), one can see that we have the following equations:\footnote{It has to be noted that one has to solve the equations in this limit exactly. This is impossible to do analytically, but  one can find the asymptotic behavior of the solution at $r\rightarrow 0$ as it is done here. In fact solving the equations numerically, shows that the obtained behaviors are correct. }
\begin{align}
& p'-p^2\frac{r_s}{r}\simeq 0  \ \ \ \ \ \textrm{as } r\rightarrow 0 \ , \\
& l'+l\frac{r_s}{r}-p\frac{r_s^2}{r^2}\simeq 0\ \ \ \ \ \textrm{as } r\rightarrow 0 \ .
\end{align}
Solving these equations we get
\begin{align}
& p=\frac{-1}{C_1+\ln(r/r_s)}\simeq \frac{-1}{\ln(r/r_s)}  \ \ \ \ \ \textrm{as } r\rightarrow 0 \ , \\
& l=C_2\frac{r_s}{r}-\frac{r_s}{r}\ln\left(\ln(r_s/r)\right)\simeq -\frac{r_s}{r}\ln\left(\ln(r_s/r)\right)\ \ \ \ \ \textrm{as } r\rightarrow 0
\end{align}
where $C_1$ and $C_2$ are constants. The metric for this region is thus
\begin{align}
& g_{00}=f(r)=1-\frac{r_s}{r}-\frac{r_s}{r}\ln\left(\ln(r_s/r)\right)  \ \ \ \ \ \textrm{as } r\rightarrow 0  \ , \label{eq4.12} \\
& g_{11}=g(r)=-\frac{1}{1-r_s/r}-  \frac{1}{\ln(r/r_s)}\ \ \ \ \ \textrm{as } r\rightarrow 0 \label{eq4.13}  \ .
\end{align}

An important property of the obtained solution is the location of the event horizon. 
The horizon(s) is(are) located at the zeros of $f(r)$ (of equation \eqref{eq4.10}). Thus, the black hole saves its classical horizon at $r_s$ or form it at
\begin{equation} \label{eq4.11}
r^{(IM)}_{H}(M_{s}) = \frac{r_{s} }{3 (1+\Omega_{M_{s}} A)} \left(1+\Omega_{M_{s}}(A+B)+\mathcal{F}_{\Omega_{M_s}}(A,B)\right) \ ,
\end{equation}
where $\mathcal{F}_{\Omega_{M_s}}(A,B) $ is a function of constants $A$ and $B$. For the classical limit $\Omega_{M_{s}}\rightarrow 0 $, this tends to the classical value, $r_{s}$.

It should be noted that, although the lapse function \eqref{eq4.10} could potentially have three roots, but two of them are avoided to be considered as a horizon because of the non--real values.

\begin{figure}
    \centering
    \begin{minipage}{0.5\textwidth}  \nonumber
        \centering
        \includegraphics[width=0.9\textwidth]{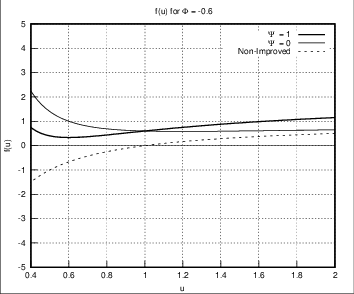}
    \end{minipage}\hfill
    \begin{minipage}{0.5\textwidth} \nonumber
        \centering
        \includegraphics[width=0.9\textwidth]{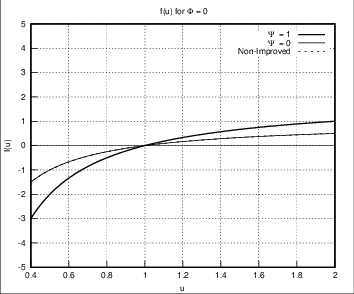} 
    \end{minipage}
    \centering
    \begin{minipage}{0.5\textwidth} \nonumber
        \centering
       \includegraphics[width=0.9\textwidth]{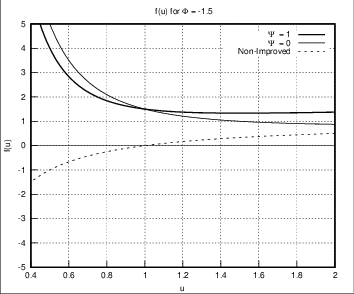}
    \end{minipage}\hfill
    \begin{minipage}{0.5\textwidth} \nonumber
        \centering
        \includegraphics[width=0.9\textwidth]{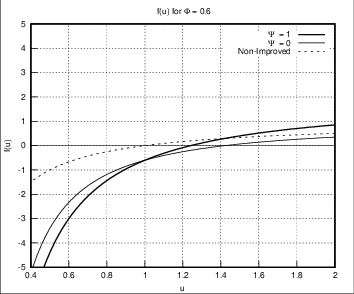}
    \end{minipage}
    \centering
   \begin{minipage}{0.5\textwidth} \nonumber
        \centering
       \includegraphics[width=0.9\textwidth]{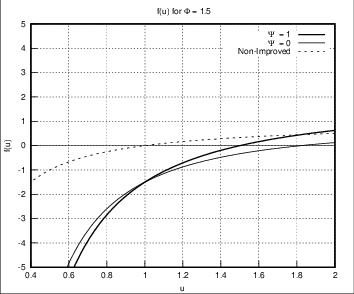}
    \end{minipage}\hfill
   \begin{minipage}{0.5\textwidth} \nonumber
        \centering
        \includegraphics[width=0.9\textwidth]{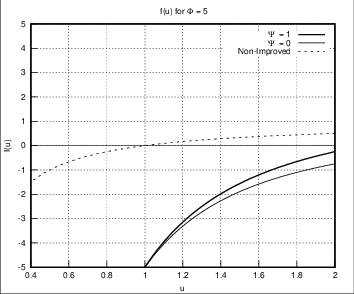}
    \end{minipage} \vspace{0.5cm}
    \caption{Improved $f(u = r/r_{s})$ for normalized quantities $\Psi = A \Omega_{M_{s}} $ and $\Phi = B \Omega_{M_{s}} $.}
    \label{Fig1}
\end{figure}

In Figure (\ref{Fig1}), the lapse function, $f(r)$ for a black hole of mass ten times the sun mass for different values of the normalized  constants $\Psi=A\Omega_{M_{s}}$ and $\Phi=B\Omega_{M_{s}}$ is plotted. It should be noted that since we are dealing with the perturbative regime with $k(r)=\xi /r$ as a cutoff identification,  this improved metric has not to be used for $r\ll r_{s}$ $(u=r/r_s\ll 1)$. From these plots, it is clear that the quantum improvement can change the location and number of horizons.

\begin{figure}
    \centering
    \begin{minipage}{0.5\textwidth}  \nonumber
        \centering
        \includegraphics[width=1\textwidth]{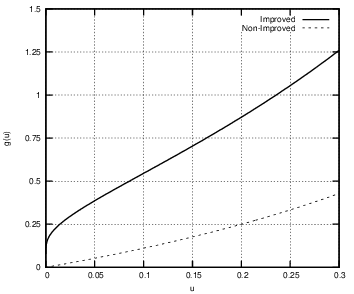}
    \end{minipage}\hfill
    \begin{minipage}{0.5\textwidth} \nonumber
        \centering
        \includegraphics[width=1\textwidth]{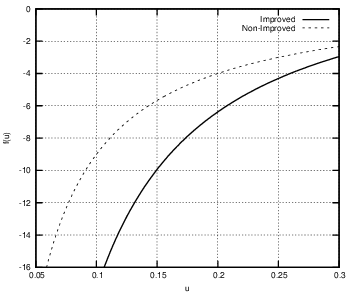} 
    \end{minipage}
        \caption{The behavior of metric components near the singularity as a function of $u=r/r_{s} $.}
    \label{addedfig}
\end{figure}

The near singularity (UV regime) behavior of the metric components is shown in Figure (\ref{addedfig}). It is interesting to note that the solution near the singularity is independent of the parameter $\Omega_{M_s}$.

\begin{figure}
    \centering
    \begin{minipage}{0.5\textwidth} \nonumber
        \centering
        \includegraphics[width=1\textwidth]{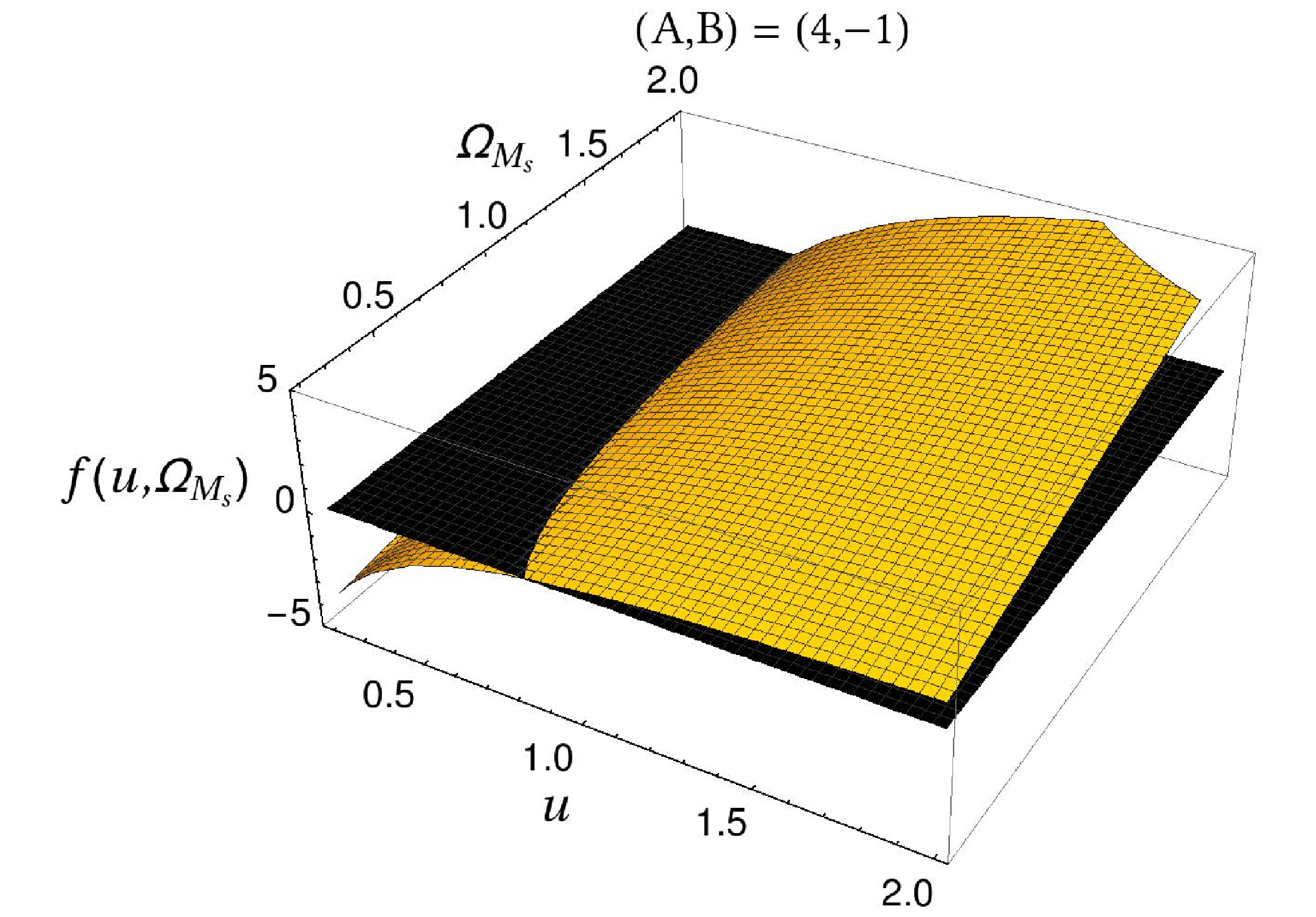}
    \end{minipage}\hfill
    \begin{minipage}{0.5\textwidth} \nonumber
        \centering
        \includegraphics[width=1\textwidth]{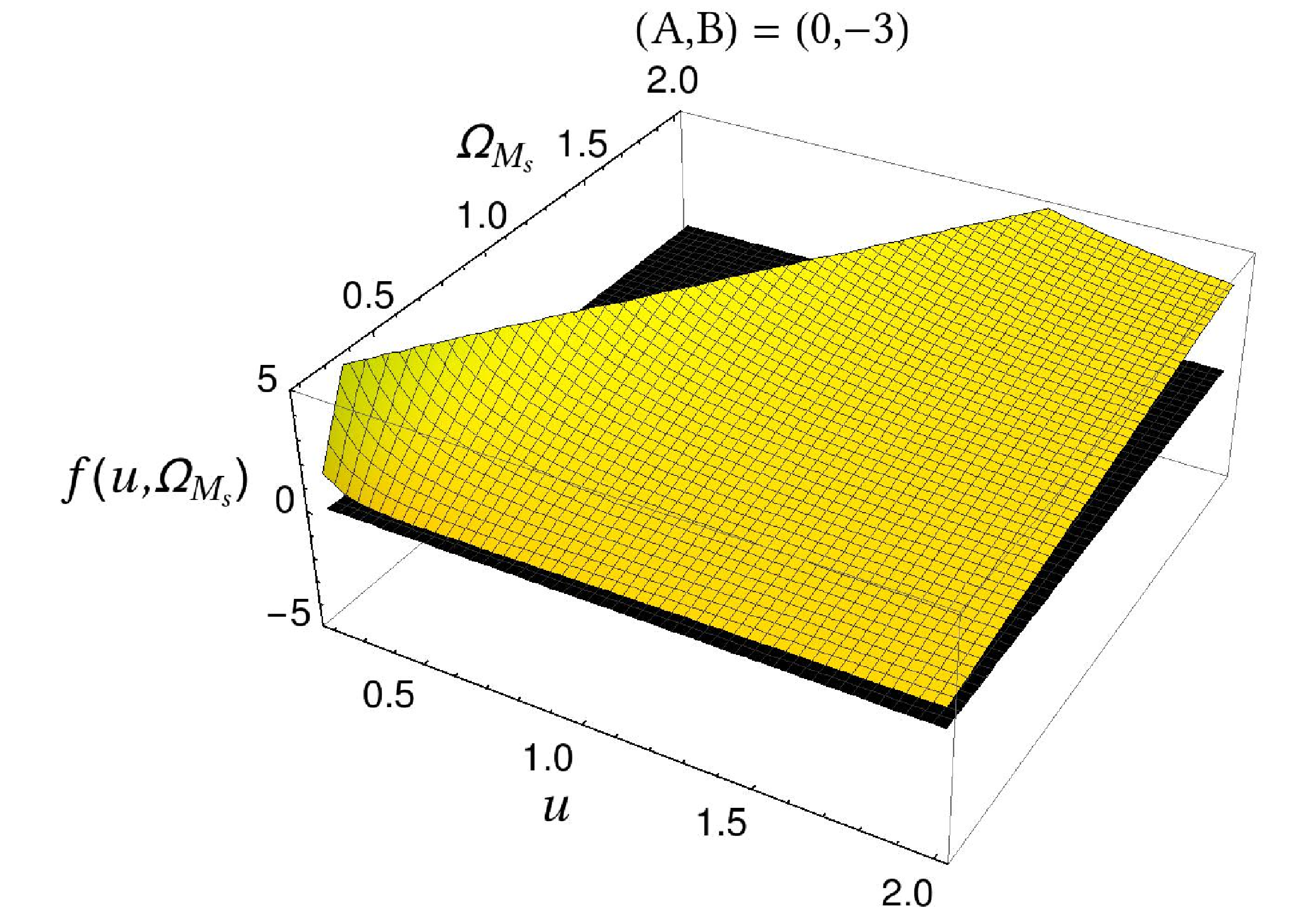}
    \end{minipage}
    \caption{The behavior of lapse function as a function of $u=r/r_{s} $ and $ \Omega_{M_{s}} $ for selected constants $A$ and $B$. }
    \label{Fig4}
\end{figure}

The Figure (\ref{Fig4}) shows the behavior of the lapse function as a function of both $u=r/r_{s} $ and $ \Omega_{M_{s}} $ for selected values of constants $A$ and $B$.  It can be seen that for special cases of $(A,B)$ there can be naked singularity.

\subsection{The trajectory of massive particles}
In order to see the effects of improvement of the black hole solution, we first investigate one of the classical test of gravity, i.e. the perihelion precession. 
It is easy to show that using the improved Schwarzschild metric \eqref{eq4.9} and \eqref{eq4.10} in the geodesic equation, the timelike worldline of a massive particle becomes
\begin{equation} \label{4.18}
  \dot{r}^{2} +\frac{h^{2}}{r^{2}} (1-\frac{r_{s}}{r}) -\frac{r_{s}}{r} - \Omega_{M_{s}} \left( \frac{s^2 r_{s}^2}{2r^2}+(1+\frac{h^{2}}{r^2})\frac{2\frac{r}{r_{s}}(B\frac{r}{r_{s}}-1)+1}{2\frac{r^3}{r_{s}^3}} \right) = -1 + s^{2} (1-\Omega_{M_{s}} A)
\end{equation}
where $ s = f(r) \dot{t} $ and  $ h = r^{2} \dot{\phi} $ are constants and thus the right hand side is also a constant of motion. The dots denote derivative respect to the proper time.  
The orbit equation for $ z = r^{-1} $,  would then be
\begin{equation}\label{eq4.19}
  z^{''} + z = \frac{r_{s}}{2 h^{2}} + \frac{3}{2} r_{s} z^{2} + \Omega_{M_{s}} \mathcal{P}_{B}(z) 
\end{equation}
where $\mathcal{P}_{B}(z) $ is the fourth order polynomial of $z$ and here prime denotes derivative respect to $\phi$.
Neglecting the $\mathcal{O}(3)$ and higher orders of $z$, this changes to
\begin{eqnarray} \label{eq4.20}
  z^{''} + \mathcal{A}z = \frac{1}{2h^{2}} r_{s}\mathcal{B} + \frac{3}{2}r_{s} \mathcal{C} z^2 
  \label{orbit}
\end{eqnarray}
where 
\begin{align} \label{eq4.21}
  & \mathcal{A} = 1 + \Omega_{M_{s}}\frac{r_{s}^{2}}{h^{2}} (1-\frac{s^{2}}{2}) \ , \nonumber \\
  & \mathcal{B} = 1 + \Omega_{M_{s}} B \ , \nonumber \\
  & \mathcal{C} = 1 + \Omega_{M_{s}}( B+\frac{r_{s}^2}{2 h^{2}}) \ .
\end{align}

The solution is simply:
\begin{equation} \label{eq4.22}
r = \frac{r_0}{1+e\cos\beta\phi} \ .
\end{equation}
This is an ellipse with perihelion precession, where $r_0$ is the mean radius and $e$ is the eccentricity of the ellipse. Inserting this solution in equations (\ref{4.18}) and (\ref{orbit}) we get
\begin{align}\label{eq4.23}
& \beta^2={\cal A}-\frac{3r_s^2}{2h^2}{\cal C}\frac{{\cal B}}{{\cal A}} \ ,\nonumber \\
& \frac{1}{r_0}=\frac{r_s}{2h^2}\frac{{\cal B}}{{\cal A}}\frac{{\cal A}-\frac{3r_s^2}{4h^2}\frac{{\cal C}{\cal B}}{{\cal A}}}{{\cal A}-\frac{3r_s^2}{2h^2}\frac{{\cal C}{\cal B}}{{\cal A}}} \ , \nonumber \\
& e^2=\frac{r_0^2}{h^2\beta^2}\left( s^2(1-\Omega_{M_s}{\cal A})-1\right) \ .
\end{align}   

In terms of physical parameters, one obtains that the perihelion advances each turn by

\begin{equation}\label{eq4.24}
\delta\phi_{\textrm{improved}}=\delta\phi_{\textrm{non--improved}}\left (1-\Omega_{M_s}\frac{r_s}{3r_0}(e^2-1)\right).
\end{equation}

The correction term is so small that has no observable result.

This is what one expects, because the quantum correction of gravity are not expected to have any important effect on planetary motions. 
\subsection{Improved black hole thermodynamics}
As a second application of the improved solution, we investigate the thermodynamics of improved black holes.
The thermodynamics of black hole in the non--improved general relativity obeys the Benkenstein--Hawking formalism which by considering the Einstein equation as a definition of state equation ends in $T = \hbar c f'(r_{H})/4\pi k_{B}$ for the temperature of the black hole's horizon, where $k_{B}$ and $\hbar$ are the Boltzman and Planck constants respectively and prime denotes differentiation respect to the coordinate $r$.

Using the same approach, for the radial improved equation of motion $ G_{r}^{r} = 8\pi G(r)(X_{r}^{r}+T_{r}^{r}) $ (where $T_{r}^{r}$ is the radial component of the perfect fluid energy--momentum tensor $T_{\mu}^{\nu}= \textrm{diag}(\rho,-P,-P,-P) $, with $\rho$ and $P$ as fluid density and pressure respectively), we would have

\begin{equation}\label{eq4.25}
  f^{'}(r)+f(r)g(r)+f(r)=-8\pi G(r)f(r)g(r)r^{2} P+8\pi G(r)(\frac{5r_s}{2r}-2) \partial_r \mathcal{J}
\end{equation}
up to first order.
Considering this equation at the horizon, $r_{H}$, along with an infinitesimally displacement of the horizon, $dr_{H}$, one gets
\begin{equation}\label{eq4.26}
 - \frac{f^{'}(r_{H})-8\pi G(r)(\frac{5r_s}{2r}-2) \partial_r \mathcal{J}}{4\pi f(r_{H}) g(r_{H})} \times \frac{2\pi r_{H}dr_{H}}{G(r_{H})} = P d(\frac{4}{3} \pi r_{H}^{3}) + \frac{dr_{H}}{2 G(r_{H})} \ .
\end{equation} 
Comparing this with the first law of thermodynamics, $ T dS = P dV + dE$, the temperature of the outer horizon of the improved black hole becomes $ T_{IM} = -\hbar c f'(r_{H})/4\pi k_{B} f(r_{H})g(r_{H}) $.
For the improved solutions \eqref{eq4.9} and \eqref{eq4.10} up to the first order, we would have
\begin{equation} \label{eq4.15}
  T_{IM} \simeq \frac{\hbar c}{4\pi k_{B}}\frac{r_{s}}{r_{H}^{2}(M_{s})} \left[ \frac{ 1 + \Omega_{M_{s}}\left(A+B+r_{s}/r_{H}(M_{s})-15r_{s}^2/4r_{H}^2(M_{s})\right)}{ 1 + \Omega_{M_{s}}\left(A - 3r_{s}^2/2r_{H}^2(M_{s})\right)}\right] \ .
\end{equation}
For $\omega \rightarrow 0 $  this tends to its non--improved value.

It has to be noted that the non--improved black hole thermodynamics is usually obtained by writing the classical Einstein's equations as the second law of thermodynamics. It is a well--known fact that doing so is equivalent to the investigation of semi--classical gravity. 
On the other hand the improved result coming from solving ERGE (used here)  contains all loop corrections \cite{Bonanno & Reuter}. As stated before the results would be in agreement with the semi--classical ones for temperatures less than the Planck temperature. 

\begin{figure}
    \centering
    \begin{minipage}{0.5\textwidth}  \nonumber
        \centering
        \includegraphics[width=0.9\textwidth]{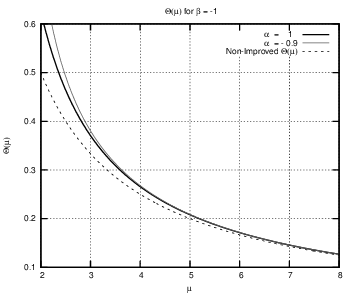}
    \end{minipage}\hfill
    \begin{minipage}{0.5\textwidth} \nonumber
        \centering
        \includegraphics[width=0.9\textwidth]{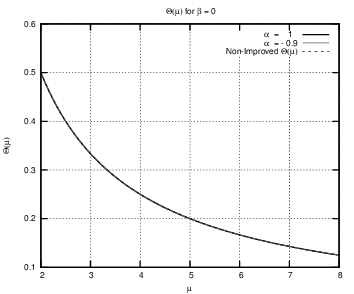} 
    \end{minipage}
    \centering
    \begin{minipage}{0.5\textwidth} \nonumber
        \centering
        \includegraphics[width=0.9\textwidth]{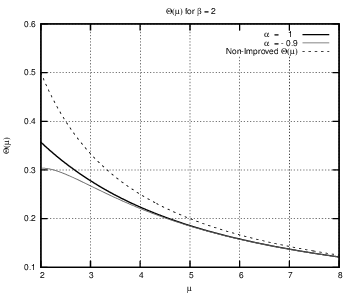}
    \end{minipage}\hfill
    \begin{minipage}{0.5\textwidth} \nonumber
        \centering
       \includegraphics[width=0.9\textwidth]{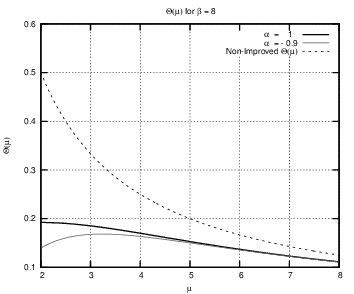}
    \end{minipage}
    \caption{The evolution of dimensionless $\Theta(\mu)$ for different values of $\alpha$ and $\beta$.} 
    \label{Fig2}
\end{figure}

By defining the dimensionless parameter $\mu \equiv M_{s}/M^{\odot}_{s}$ where $ M^{\odot}_{s} $ is the Schwarzschild mass of sun, we would have
\[
r_{s} = r_{s}^{\odot} \mu  \quad , \quad r_{H} = r_{s}^{\odot}\bar{r}_{H}(\mu) \quad,\quad T_{IM} = T_{0}^{\odot} \Theta(\mu)
\]
where $ r_{s}^{\odot} = 2 G M_{s}^{\odot}/c^{2} $ is the Schwarzschild radius of sun, $T_{0}^{\odot} = \hbar c/4\pi k_{B}r_{s}^{\odot}$ and
\begin{align}
  & \bar{r}_{H}(\mu) = \frac{\mu}{3(\alpha+\mu^2)} \left(\alpha+\beta +\mu^{2}+ \mu^{2} \mathcal{F}_{\Omega_{M_s}}(A,B)\right) \ , \\
  & \Theta(\mu) = \frac{\mu}{\bar{r}_{H}^{2}(\mu)} \frac{\mu^{2}+\alpha+\beta+\Omega_{M_{s}^{\odot}}(\frac{\mu}{\bar{r}_{H}(\mu)}-\frac{15\mu^2}{4\bar{r}_{H}^2(\mu)})}{\mu^{2}+\alpha-\Omega_{M_{s}^{\odot}}\frac{3\mu^{2}}{2\bar{r}_{H}^{2}(\mu)}}
\end{align}
are the dimensionless improved horizon and temperature, respectively.  We also used the  renaming $\alpha =\Psi^\odot= A \Omega_{M_{s}^{\odot}} $ and $\beta =\Phi^\odot= B \Omega_{M_{s}^{\odot}} $. The dependence of $\Theta$ on $\mu$ is drawn in Figure (\ref{Fig2}).

\begin{figure}
    \centering
    \begin{minipage}{0.5\textwidth}  \nonumber
        \centering
        \includegraphics[width=0.9\textwidth]{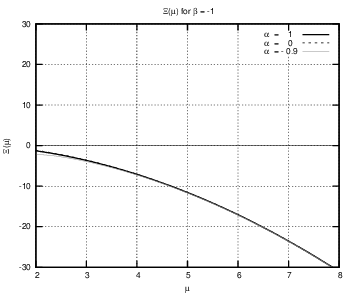}
    \end{minipage}\hfill
    \begin{minipage}{0.5\textwidth} \nonumber
        \centering
        \includegraphics[width=0.9\textwidth]{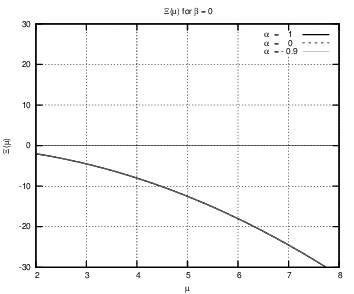} 
    \end{minipage}
    \centering
    \begin{minipage}{0.5\textwidth} \nonumber
        \centering
        \includegraphics[width=0.9\textwidth]{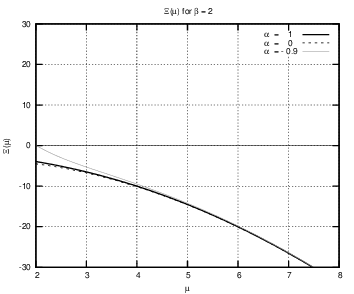}
    \end{minipage}\hfill
    \begin{minipage}{0.5\textwidth} \nonumber
        \centering
       \includegraphics[width=0.9\textwidth]{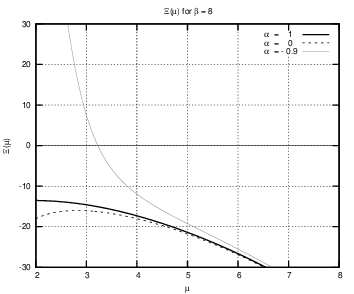}
    \end{minipage}
    \caption{Normalized heat capacity $\Xi(\mu)$.} 
    \label{Fig3}
\end{figure}

Using the improved definition $ dE = \frac{dr_{H}}{2G(r_{H})} $ (see equation (\ref{eq4.26})), the heat capacity $ C(M_{s}) = \frac{\partial E}{\partial M_{s}} \frac{\partial M_{s}}{\partial T} $ becomes
\begin{equation} \label{eq4.17}
   C(M_{s}) \equiv C_{0} \Xi(\mu)
\end{equation}
where  $C_{0}=r_{s}^{\odot}/G_{N}T_{0}^{\odot}$ and $ \Xi(\mu) = \frac{1}{2\bar{G}(\mu)}\frac{d \bar{r}_{H}}{d\mu} \frac{d\mu}{d\Theta}$ is the dimensionless heat capacity with $\bar{G}(\mu) = 1- \Omega_{M_{s}^{\odot}}/6\bar{r}_{H}^{2}(\mu)$.
The $\mu$-dependence of heat capacity is plotted in Figure (\ref{Fig3}).

From figures (\ref{Fig2}) and (\ref{Fig3}), it is clear that for negative (positive) values of $\beta$ the effect of the quantum improvement is raising (lowering) the black hole temperature and lowering (raising) the specific heat for each value of mass.

\section{Discussion and concluding remarks}\label{4}
Although applying the RG improvement method leads to some quantum correction to general relativity, the method depends on the way one improves the classical results. Therefore
comparing the results of different methods of quantum improvement would be considerable.
Here we suggested to use some curvature invariant like ($\chi\equiv R_{\mu\nu} R^{\mu\nu}$), in order to scale the running coupling.
This substitution in the action saves the general covariance as well as leading to a self consistent set of equations. 

As it is discussed in \cite{Moti-Shojai3}, this is a better approach than parameter, equation of motion or solution improvements studied before \cite{Falls, Pawlowski, Varing G,Reuter & Weyer & Noorbala, Dittrich & Reuter}.

To investigate the results of this method, we studied the static vacuum solution of Einstein equation, i.e. Schwarzschild black hole.
To know how this method differs from others, the iterative solutions \eqref{eq4.7} and \eqref{eq4.8} should be compared with $f(r)=-g(r)^{-1} = 1 - r_{s}/r ( 1 + \frac{\Omega_{M}}{2} \frac{r_{s}^{2}}{r^{2}})$  obtained from the improved equation method (equivalent to improved vacuum solution) \cite{Bonanno & Reuter}.

Although the classical Schwarzschild metric is a vacuum solution and hence describes a spacetime with zero scalar curvature, the improvement can lead to a non--zero one.  At UV--regime where \eqref{eq4.12} and \eqref{eq4.13} are valid, the leading term of the action improved scalar curvature can simply calculated as
\[
R^{\text{A-I}}\ \stackrel{r\rightarrow 0}{\sim}\ \frac{1}{r^3}
\]
while equation improved scalar curvature behaves as \cite{Bonanno & Reuter}
\[
R^{\text{EoM-I}}\ \stackrel{r\rightarrow 0}{\sim}\ \frac{1}{r} \ .
\]
This means that the action improved scalar curvature diverges faster than the equation improved one.

\begin{figure}
        \centering
        \includegraphics[width=0.5\textwidth]{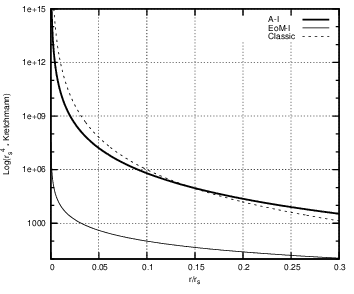}
        \caption{Near--singularity behavior of Kretschmann invariant for the classical Schwarzschild solution (dotted line), the action--improved solution (thick line), and the equation--improved one (thick line).}
        \label{Fig23}
\end{figure}

Even if the Kretschmann scalar is chosen for the cutoff identification, $k\varpropto (R_{\alpha\beta\gamma\delta}R^{\alpha\beta\gamma\delta})^{-1/4}$  \cite{Pawlowski}, the $R^{\text{EoM-I}}\varpropto r^{-3/2}$ diverges slower than $R^{\text{A-I}}$.
A better comparison can be done by looking at the behavior of the Kretschmann invariant. The $r^{-6}$ behavior of Kretschmann scalar of classical solution is \textit{nearly} preserved by the UV--regime action--improved solution. But the equation--improved result predicts a leading term of the form $r^{-2}$ and $r^{-3}$ for ($k\varpropto r^{-1}$) and ($k\varpropto (R_{\alpha\beta\gamma\delta}R^{\alpha\beta\gamma\delta})^{-1/4}$) models, respectively \cite{Bonanno & Reuter, Pawlowski}. In brief, action--improved solution is closer to the classical solution and is more singular in Kretschmann invariant than the equation--improved one. Figure \ref{Fig23} shows the  Kretschmann invariant behavior near the singularity for the three cases.

\begin{figure}
    \centering
    \begin{minipage}{0.5\textwidth}  \nonumber
        \centering
        \includegraphics[width=0.9\textwidth]{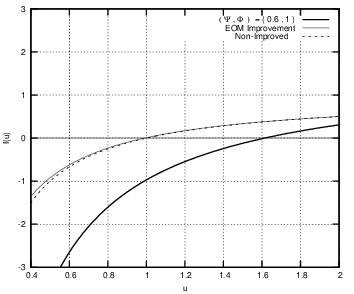}
        \caption*{(a) $\Omega_{M_{s}} \simeq \Omega_{M_{s}^{\oplus}}$}
    \end{minipage}\hfill
    \begin{minipage}{0.5\textwidth} \nonumber
        \centering
        \includegraphics[width=0.9\textwidth]{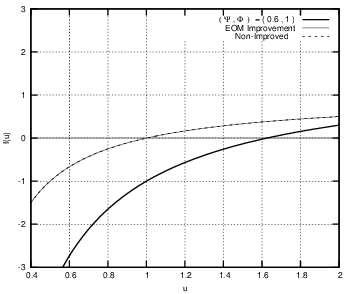} 
        \caption*{(b) $\Omega_{M_{s}} \simeq \Omega_{M_{s}^{\odot}}$}
    \end{minipage}
    \caption{Comparison of the lapse function for action--improved, equation of motion--improved and the non--improved cases. The plots are for (a) $\Omega_{M_{s}} \simeq \Omega_{M_{s}^{\oplus}}$ and (b) $\Omega_{M_{s}} \simeq \Omega_{M_{s}^{\odot}}$, where $M_{s}^{\oplus}$ and $M_{s}^{\odot}$ denote the Schwarzschild mass of earth and sun, respectively. } 
    \label{Fig5}
\end{figure}

In contrast to the results of equation of motion improvement, the formation of second horizon is impossible here. Hence, the solution of improved equation, would result in one horizon or naked singularity for this method.
\footnote{In \cite{Pawlowski}, it is claimed the choice of $ k\varpropto (R_{\alpha\beta\gamma\delta}R^{\alpha\beta\gamma\delta})^{-1/4} $ to improve the equation of motion saves the cosmic censorship conjecture, which is a notable issue.The simple comparison can be seen in Figure (\ref{Fig5}). }
It shows a simple comparison between the lapse functions of these two improvement methods for different values of $\Omega_{M_{s}}$.
It is clear that for larger masses the difference between two methods is larger.

As a check of not destroying the known results of general relativity, after determining the lapse function, we studied the massive particles' trajectory and perihelion precession. It is shown that the corrections are too small to make any observable change to the general relativity's predictions.

The thermodynamics of improved spacetime is also studied. Although the dependence of temperature on the Schwarzschild mass differs from what general relativity predicts, except for having a local maximum for some specific values of $A$ and $B$, its descending behavior does not change (see Figure (\ref{Fig2})). This is in contrast to the result  of equation of motion improvement method (for both ($k\varpropto r^{-1}$) and ($k\varpropto (R_{\alpha\beta\gamma\delta}R^{\alpha\beta\gamma\delta})^{-1/4}$) models), in which the  temperature reaches a global maximum after a sharp increase, and then starts to decrease along with general relativity's predictions \cite{Bonanno & Reuter}.

The heat capacity of black hole, $C(M_{s})$, is also a prominent thermodynamical feature. It can be seen in Figure (\ref{Fig3}) that although for some choices of $A$ and $B$ we can have positive heat capacity, but the decreasing behavior of $C(M_{s})$ remains valid. This is not compatible with the results of equation of motion (or solution) improvement method \cite{Bonanno & Reuter}.

In summary, the results of the suggested action improvement method differs in details from other methods. And it gives a more physical approach, because the improvement is not introduced as the running coupling in the solutions or in the equations of motion. It is introduced in a dynamical way such that the general covariance holds.

\vglue1cm
\textbf{Acknowledgment:} This work is supported by a grant from Iran National Science Foundation (INSF).


\begin{thebibliography}{}
  \bibitem{Moti-Shojai3} R. Moti and A. Shojai, \textit{Phys. Lett. B} \textbf{793}, 313 (2019).
  \bibitem{Time} E. Anderson, arXiv: gr-qc/1009.215.
  \bibitem{Rovelli} C. Rovelli, \textit{Qunautm Gravity} (Cambridge Monographs on Mathematical Physics), Cambridge: Cambridge University Press (2008).
  \bibitem{Weinberg-1st} S. Weinberg, in \textit{General Relativity}, edited by S. W. Hawking and W. Isreal, Cambridge: Cambridge University Press (1979).
  \bibitem{Weinberg-2nd} S. Weinberg, in \textit{Understanding the Fundamental Constituents of Matter}, edited by A. Zichichi, New York: Plenum Press (1978).
  \bibitem{Attentions} O. Lauscher and M. Reuter, \textit{Phys. Rev. D} \textbf{65}, 025013 (2002), \\
  D. F. Litim,  \textit{Phys. Rev. Lett.} \textbf{92}, 201301 (2006), \\
  A. Bonanno and M. Reuter, \textit{JHEP} \textbf{0502}, 035 (2005), \\
  A. Codello and R. Percacci, \textit{Phys. Rev. Lett.} \textbf{97}, 221301 (2006), \\
  F. Girelli, S. Liberati, R. Percacci and C. Rahmede,  \textit{Class. Quant. Grav.} \textbf{24}, 3995-4008 (2007), \\
  A. Eichhorn, \textit{Phys. Rev. D} \textbf{87}, 124016 (2013), \\
  A. Nink and M. Reuter, \textit{JHEP} \textbf{1301}, 062 (2013), \\
  M. Reuter and G.M. Schollmeyer, \textit{Ann. Phys.} \textbf{367}, 125 (2016), \\
  C. Pagani and M. Reuter, \textit{JHEP} \textbf{1807}, 039 (2018), \\
  A. Eichhorn and P.V. Griend, \textit{JHEP} \textbf{1808}, 147 (2018).
  \bibitem{Weinberg inflation} S. Weinberg, \textit{Phys. Rev. D} \textbf{81}, 083535 (2010).
  \bibitem{Reuter-1st} M. Reuter, \textit{Phys. Rev. D} \textbf{57}, 971 (1998), \\
   M. Reuter, arXiv:hep-th/9605030.
   \bibitem{Souma} W. Souma, \textit{Prog. Theor. Phys.} \textbf{102}, 181 (1999).
  \bibitem{Moti-Shojai} 
   A. Bonanno and M. Reuter,  \textit{Int. J. Mod. Phys. D} \textbf{13}, 107 (2004), \\
  A. Bonanno and  F. Saueressig,  \textit{Comptes Rendus Physique} \textbf{18}, 254 (2017), \\
  R. Moti and A. Shojai, \textit{Eur. Phys. J. C} \textbf{78}, 32 (2018), \\
  A. Platania, \textit{Universe 5} \textbf{8}, 189 (2019).
  \bibitem{Reuter Becker} M. Reuter and E. Tuiran, \textit{Phys. Rev. D} \textbf{83}, 044041 (2011),\\
 D. Becker and M. Reuter, \textit{JHEP} \textbf{1207}, 172 (2012),\\ 
 B. Koch and  F. Saueressig,  \textit{Int. J. Mod. Phys. A} \textbf{29},  1430011 (2014), \\
 A. Adeifeoba, A. Eichhorn and A. Platania, \textit{Class. Quant. Grav.} \textbf{35}, 225007 (2018).
  \bibitem{Falls} K. Falls, F. Litim and A. Raghuraman, \textit{Int. J. Mod. Phys. A} \textbf{27}, 1250019 (2012).
 \bibitem{Bonanno & Reuter} A. Bonanno and M. Reuter, \textit{Phys. Rev. D} \textbf{62}, 043008 (2000), \\
   A. Bonanno and M. Reuter, \textit{Phys. Rev. D} \textbf{65}, 043508 (2002).
  \bibitem{Pawlowski} J. M. Pawlowski and D. Stock, \textit{Phys. Rev. D} \textbf{98}, 106008 (2018).
  \bibitem{Varing G} V. Faraoni, \textit{Cosmology in Scalar--Tensor Gravity} (Fundamental theories of Physics), Netherlands: Kluwer Academic Publishers (2004).
  \bibitem{Reuter & Weyer & Noorbala} M. Reuter and H. Weyer, \textit{Phys. Rev. D} \textbf{69}, 104022 (2004).
  \bibitem{Wetterich} C. Wetterich, \textit{Phys. Lett. B} \textbf{301}, 90-94 (1993).
  \bibitem{Nagy} S. Nagy, \textit{Ann. Phys.} \textbf{350}, 310 (2014).
  \bibitem{Dittrich & Reuter} W. Dittrich and M. Reuter, \textit{Effective Lagrangians in Quantum Electrodynamics}, Berlin: Springer (1985).
  \bibitem{Ellis} G. F. R. Ellis and J. Uzan, \textit{Am. J. Phys.} \textbf{73}, 240 (2005).
  \bibitem{Moti-Shojai4} R. Moti and A. Shojai,  arXiv:1909.07959.
\end{thebibliography}
\end{document}